\documentclass[twocolumn,aps,superscriptaddress]{revtex4}

\usepackage{amsmath}
\usepackage{graphicx}

\usepackage{float}
\floatstyle{ruled}
\newfloat{floatbox}{htp}{box}
\floatname{floatbox}{Box}
\newcommand{\infobox}[2]{
    \begin{floatbox}
        \caption{#1}
        \let\centering\relax
        \parbox{.94\columnwidth}{#2}
    \end{floatbox}
}

\newcommand{\ket}[1]{\vert #1 \rangle}

\begin{document}

\title{Hybrid Quantum Information Processing}

\author{Ulrik L. Andersen}
\email[Corresponding author:]{ulrik.andersen@fysik.dtu.dk}%
\author{Jonas S. Neergaard-Nielsen}
\affiliation{Department of Physics, Technical University of Denmark, Fysikvej, 2800 Kongens Lyngby, Denmark}
\author{Peter van Loock}
\affiliation{Institute of Physics, Johannes-Gutenberg Universit\"at Mainz, Staudingerweg 7, 55128 Mainz, Germany}
\author{Akira Furusawa}
\affiliation{Department of Applied Physics, School of Engineering, The University of Tokyo,
7-3-1 Hongo, Bunkyo-ku, Tokyo 113-8656, Japan}

\maketitle

{\bf The development of quantum information processing has traditionally followed two separate and not
immediately connected lines of study. The main line has focused on the implementation of quantum bit (qubit) based
protocols whereas the other line has been devoted to implementations based on high-dimensional Gaussian states (such as
coherent and squeezed states). The separation has been driven by the experimental difficulty in interconnecting the
standard technologies of the two lines.  However, in recent years, there has been a significant experimental progress
in refining and connecting the technologies of the two fields which has resulted in the development and experimental
 realization of numerous new hybrid protocols. In this Review, we summarize these recent efforts on hybridizing the two types of schemes based on discrete and continuous variables.}

\section{Introduction}
By harnessing the quantum effects of superposition
and entanglement, revolutionary new methods of communication and computation can be realized~\cite{Nielsen2000}. A
prominent example is the possibility of generating a secret key between two or more parties
in a communication network leading to unconditionally secure
communication -- known as quantum cryptography. Another striking example, and perhaps the most
intriguing one, is the realization of a quantum computer which allows for exponentially
faster computation of certain tasks. Although the experimental progress
in controlling quantum states of various microscopic
quantum systems has exploded in recent years, the
implementation of a fully fault-tolerant and scalable quantum computer is
still a major challenge.

Spurred by the grand vision of constructing a quantum computer, numerous physical platforms are being intensely
explored worldwide. These include light, ions, atoms, solid state, cavity quantum electrodynamics, superconducting systems and nuclear magnetic resonance~\cite{Ladd2010}. 
However, irrespective of the physical system, quantum information processing (QIP) comes in two different forms
depending on which degree of freedom, or observable, is being used for describing the
involved quantum states. If this observable is of discrete nature (that is, its eigenvalues are discretized), one often
refers to discrete-variable (DV) QIP (Box 1), and if the observable has a continuum of eigenvalues, one talks about continuous-variable (CV) QIP (Box 2). An analog can be drawn to classical information processing where the two forms exist in the realm of digital and analog information processing.

In recent years, numerous research groups have worked on bridging the two islands with the aim of realizing protocols
that overcome the intrinsic limitations of the individual DV and CV schemes. The integration of DV and CV technologies
in unified hybrid optical systems has thus resulted in a series of new proposals and ground-breaking experiments
realizing long-standing goals~\cite{VanLoock2011}. The aim of this Review is to present the recent progress on
combining the CV and DV methods for applications in QIP. A major research effort has been devoted to the
generation of highly nonclassical quantum states using hybrid technologies. These efforts will be reviewed in section
\ref{sec:nongauss_generation}. In section \ref{sec:qip}, we will review QIP including quantum teleportation, quantum repeaters and quantum computing
based on hybrid schemes, and finally, in section \ref{sec:outlook}, we conclude the review with an outlook.

\infobox{Discrete-variable quantum information processing} {A single
photon can carry information in different degrees of freedom such as its
polarization, its time of arrival and spatial mode whereas an electron can carry information in its spins. 
Binary digits can
thus be represented by orthogonal eigenstates of a single photon or a single electron. In
quantum information, the information is not solely comprised by the
eigenstates, but can be encoded as superpositions of the eigenstates:
\begin{equation}
\label{eq:qubit}
|\psi\rangle=c_0|0\rangle+c_1|1\rangle
\end{equation}
which is known as the qubit. The information of this quantum state is
given by the complex amplitudes $c_0$ and $c_1$, and it can be visualized
on the Bloch sphere. In this system the computational basis set $\{|0\rangle,|1\rangle\}$
is discrete and limited to two. The measurement of this qubit is described
by a two-component projector
such that in each measurement the number of outcomes (eigenvalues) is
limited to two. An example is a Stokes parameters measurement (consisting
of waveplates, a polarizing beam splitter and two photon counters) or a Stern-Gerlach apparatus both of which
ideally projects along any orthogonal basis. A universal two-component projector can be used to implement a measurement induced
non-linearity and it can be used to fully characterize a state in the two-dimensional Hilbert space.

In order to implement universal quantum computation and perform a complete set of quantum communication tasks, a
finite
set of gates comprising single qubit and two-qubit operations must be
implemented. One example of a complete set is $\{\hat U_{H},\hat U_{PG},\hat U_{CNOT}\}$
where $\hat U_H$ and $\hat U_{PG}$ are the single qubit Hadamard and rotation gates and $\hat U_{CNOT}$ is the two
qubit controlled NOT gate. Deterministic single qubit gates can be
readily performed with simple linear optics but the deterministic
two-mode gate requires the introduction of a very large non-linearity. These hard interactions can however be by-passed using probabilistic measurement induced operations as suggested in ref. \cite{Knill2001} but the required overhead is massive \cite{Kok2007}.

\vspace{.5em}
\includegraphics[width=.8\columnwidth]{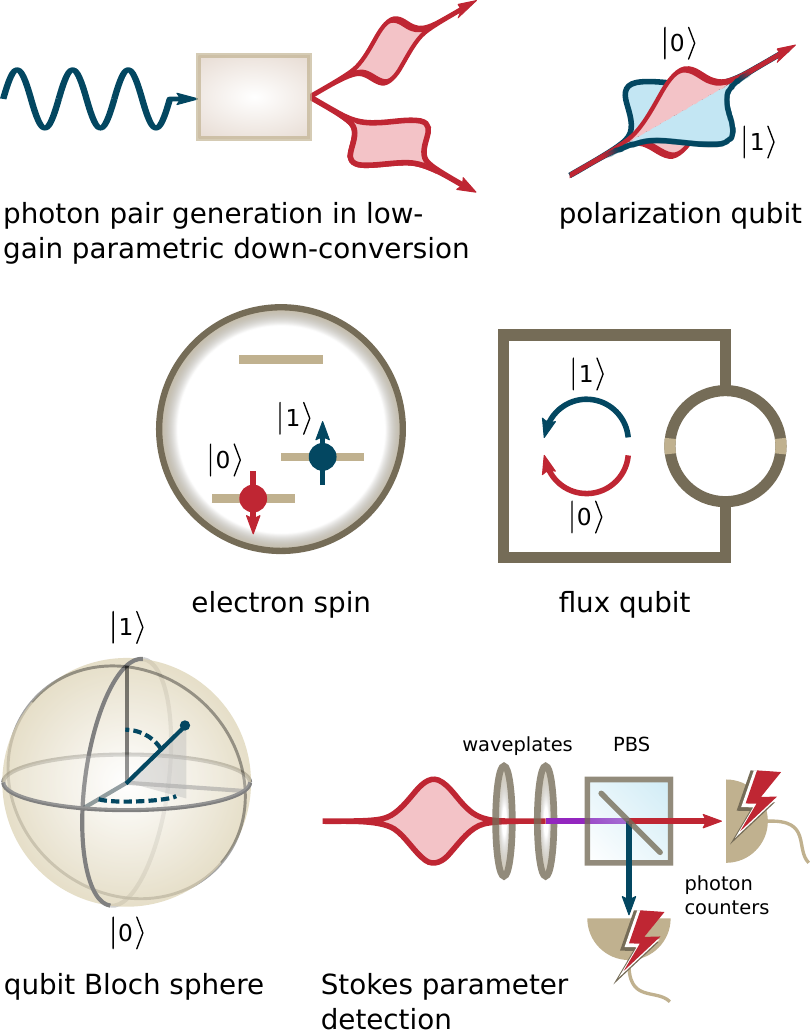}

}

\infobox{Continuous-variable quantum information processing} {
As an alternative to the standard finite-level encoding, one
might use a continuous basis $\{|x\rangle\}$ for the encoding, examples being the amplitude and phase quadratures of a field mode, the postion and momentum of a mechanical oscillator and the spin variables of an atomic ensemble.  
An arbitrary quantum state in this basis is
\begin{eqnarray}
|\psi\rangle=\int{\psi(x)|x\rangle dx}
\end{eqnarray}
where the information is now contained in the wavefunction $\psi(x)$ rather than in discrete numbers as for
the two-level system (see Box 1). If $\psi(x)$ is Gaussian, the state is coined Gaussian which is the case for the coherent state, the squeezed state and the CV entangled state (also known as the two-mode squeezed state). In optics, the position eigenstates
correspond to a quadrature eigenstate which physically represents an
infinitely squeezed state displaced to the quadrature value $x$. A measurement of the
basis states is done with a continuous projector which physically can be carried out with high efficiency using a
homodyne detector. The outcomes of such measurements, that is, the
eigenvalues, are now continuous numbers. Using such a homodyne projector, it is possible to perform complete tomography
of any quantum state of light.

A universal set of gates for
continuous-variable computation has been defined and they can basically
be categorized in two types of transformations; Gaussian and non-Gaussian
transformations: $\{ \hat F,\hat Z,\hat U_{SUM},\hat U_{PG}\}$ including the single mode Gaussian gates (Fourier
transform and displacement), the g Gaussian SUM gate and the two-mode non-Gaussian cubic phase gate. The
Gaussian transformations are standard in a CV laboratory and sufficient for a
large range of protocols ~\cite{Furusawa1998,Grosshans2003,Aoki2009,Lassen2010}. However, universality is only attained by the
technically challenging non-Gaussian transformations~\cite{Bartlett2002,Fiurasek2002,Eisert2002,Giedke2002,Niset2009}.

\vspace{.5em}
\includegraphics[width=.8\columnwidth]{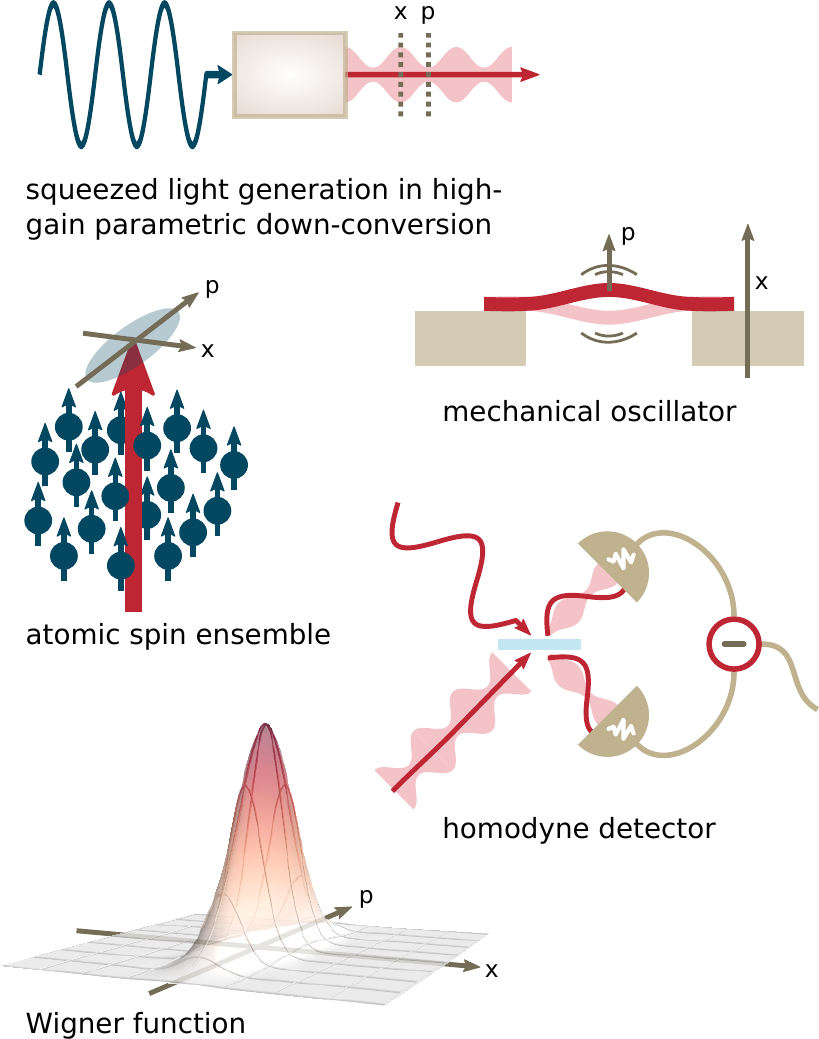}
}

\section{Generation of non-Gaussian states}
\label{sec:nongauss_generation}

\infobox{The Wigner function}
{
The quantum state of an optical field
can be represented as a distribution $W(x,p)$ over the
continuous-variable phase space. This \textsl{Wigner function} has a
one-to-one correspondence with the density matrix, and therefore
contains all information about the state. It is very similar to a
classical probability distribution, but it can attain negative values.
This property, closely related to Heisenberg's uncertainty principle,
provides a straight-forward criterium for the non-classicality of a
quantum state. Among pure states, only the Gaussian ones do not exhibit
negativity of their Wigner functions. But negativity is fragile, so it
can quickly vanish from a non-Gaussian state that has lost its purity.
Observation of non-Gaussian states with negative Wigner functions
therefore require high efficiencies throughout the system. The Wigner
function can be reconstructed by the method of homodyne tomography,
where multiple identical copies of the quantum state are prepared and
measured on a homodyne detector with different settings of the phase of
the local oscillator. Alternatively, it can reconstructed via displacement controlled parity measurement.
\vspace{1ex}
}

There are two classes of pure quantum states that play a pivotal role in quantum information processing: Gaussian states and non-Gaussian states, referring to the statistics of the state's wavefunction or Wigner function (Box 3). The Gaussian states -- e.g. coherent and squeezed states -- are often referred to as CV states. They are relatively easy to produce and manipulate using standard CV technology such as lasers, parametric amplifiers (or squeezers), beam splitters and homodyne detectors. This technology enables a linear transformation of continuous quantum quadratures thereby mapping a Gaussian state onto another Gaussian state \cite{Braunstein2005,Andersen2010,Weedbrook2012}. 
Such transformations have been mastered in the optical regime for more than two decades, but in recent years it has been also extended to the microwave regime: Using superconducting degenerate \cite{Castellanos-Beltran2008} and non-degenerate \cite{Bergeal2012,Eichler2011a} parametric amplifiers, microwave squeezed and CV entangled states have been generated and characterized with homodyne detection for state tomography \cite{Eichler2012}. In addition to the generation of squeezing of the field quadratures, there has also been some recent demonstrations on squeezing the CV collective spin observables of an atomic ensemble \cite{Esteve2008} and similar proposals exist for solid state materials \cite{Rudner2011,Bennett2013}. 
Within the last decade, there has been massive interest in generating and manipulating the position and momentum CVs of mechanical oscillators. This has lead to numerous proposals on generating mechanically squeezed and entangled states exploiting the Gaussian coupling between a field mode and the mechanics \cite{Aspelmeyer2013}. This has very recently been accomplished in a superconducting system in which entanglement between a microwave field and a mechanical oscillator was created \cite{Palomaki2013}.    
 
To produce pure non-Gaussian states, and in general an arbitrary quantum state, the standard CV toolbox consisting of linear Gaussian transformation and homodyne detection is insufficient. It is however possible to enter the non-Gaussian regime by hybridizing DV and CV technology. There are basically two approaches to the formation of non-Gaussian states of an oscillator: 1) By enabling a strong, deterministic coupling to a finite-level (discretized) matter system or 2) by a probabilistic measurement-induced interaction using a finite-level discretized energy detector (photon counter). 
  
{\bf Deterministic generation of non-Gaussian states using two-level matter systems.} The interaction between a CV oscillator and a DV two-level system can be described by the Jaynes-Cummings interaction. 
The simplest non-Gaussian state to produce using such an interaction is the single photon state; each time the two-level system is excited, it will decay and emit a single flying photon into a travelling field mode. If a single field mode is strongly coupled to the two-level system -- usually enabled by placing the systems inside a high Q cavity -- the photon will be harvested by that mode with large probability. This is known as the Purcell effect. It has been demonstrated in a number of experiments~\cite{Eisaman2011}, but a complete state characterization via Wigner function reconstruction has been realized only in a few experiments, mainly in the microwave regime \cite{Deleglise2008,Eichler2011b,Mallet2011,Menzel2010}, but recently also in the optical regime with atomic ensembles~\cite{MacRae2011,Bimbard2014}.

\begin{figure*}[htp]
  \includegraphics[width=.8\textwidth]{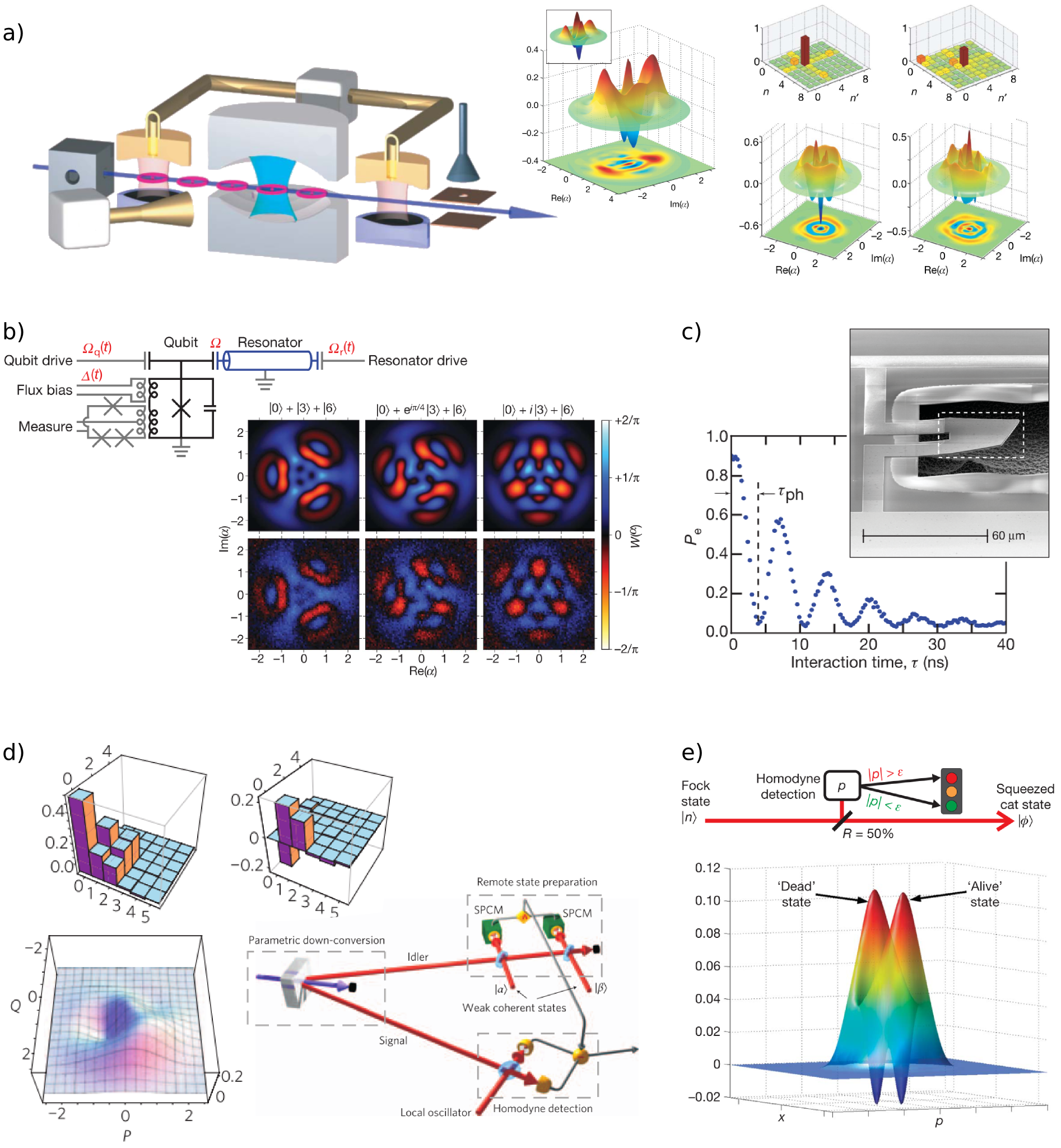}
  \caption{
    Examples of non-Gaussian state generation in various systems. 
    \textbf{a)} Schr\"odinger cat and Fock states of a microwave cavity field induced by the detection of dispersively coupled Rydberg atoms \cite{Deleglise2008},
    \textbf{b)} arbitrary Fock state superpositions in a waveguide resonator field coupled to a superconducting phase qubit \cite{Hofheinz2009},
    \textbf{c)} population exchange of a single excitation between a superconducting phase qubit and a piezoelectric mechanical oscillator cooled to its ground state \cite{O'Connell2010},
    \textbf{d)} arbitrary Fock state superpositions of an optical mode through spontaneous parametric down-conversion and coherent-state injected photon detectors \cite{Bimbard2010},
    \textbf{e)} squeezed Schr\"odinger cat state of an optical mode induced by conditional homodyne detection on a 2-photon Fock state \cite{Ourjoumtsev2007a}.
  }
  \label{fig:systems}
\end{figure*}

In the microwave domain, the coupling strength can be enormously high by employing a superconducting phase qubit inside a microstrip cavity, and moreover, the coupling can be controlled by  detuning the cavity in and out of resonance with the field~\cite{Law1996}. Using such a strong and controlled coupling, higher order Fock states \cite{Hofheinz2008} as well as Fock state superpositions of several photons \cite{Hofheinz2009} have been deterministically generated and characterized by quantum state tomography on a chip in a cryogenic environment, see Fig. \ref{fig:systems}b. Similar multi-photon Fock states have been generated via quantum non-demolition measurement of a stationary microwave field employing the strong coupling and subsequent measurements of flying Rydberg atoms \cite{Deleglise2008}, see Fig. \ref{fig:systems}a.   

Another non-Gaussian state of the harmonic oscillator that has attracted significant interest is the so-called ``Schr\"odinger cat state''. It is defined as a superposition of macroscopic states. Examples are the superpositions of
coherent states with opposite phase, $\ket{\alpha} +e^{i\phi}\ket{-\alpha}$ \cite{Yurke1986} where $\alpha$ is the coherent state amplitude and $\phi$ is a phase. The name is a reference to
Schr\"odinger's famous Gedankenexperiment \cite{Schrodinger1935}: The coherent states are seen
as macroscopically distinct states, ``alive'' and ``dead,'' and the cat
state is in a superposition of these. 
Such states have been deterministically realized in the motional state of a trapped ion \cite{Monroe1996} and in the microwave regimes by entangling a standing CV microwave field to a flying Rydberg atom followed by a projective DV measurement of the atom \cite{Brune1996,Deleglise2008} and through a strong, dispersive interaction with a superconducting transmon qubit \cite{Vlastakis2013}. 
 
In addition to the coupling of two-level systems to the CVs of the electro-magnetic field, recently there has also been significant progress in understanding and implementing the coupling of a two-level system to the continuous position and momentum variables of a mechanical oscillator. In a pioneering experiment it was shown that by strongly coupling a superconducting phase qubit to a ground state cooled mechanical oscillator, it was possible to generate a single excitation (single phonon Fock state) of the oscillator \cite{O'Connell2010}, see Fig. \ref{fig:systems}c. 
Various other approaches enabling a strong coupling of a two-level system (e.g. a quantum dot \cite{Wilson-Rae2004}, an NV center in diamond \cite{Rabl2009}, an atom \cite{Hammerer2009} and a two-level defect \cite{Ramos2013}) to a mechanical oscillator have been proposed and some recent preliminary steps have been realized \cite{Bennett2010,Yeo2014,Arcizet2011,Kolkowitz2012}. These schemes promise the formation of mechanical oscillators in arbitrary superposition states including the cat state, which in principle can be mapped onto the electromagnetic field  \cite{Stannigel2010}. This might well be the future route to determinisitic generation of non-Gaussian states for QIP.

{\bf Probabilistic generation of non-Gaussian states using finite-level detection systems.}
In the optical regime, the non-Gaussian transformation is notoriously difficult to implement due to the non-existence of sufficiently strong, optical non-linearities. 
As an alternative to the deterministic scheme, one may generate the transformation probabilistically using a measurement-induced nonlinearity by applying a 
non-Gaussian projector. The simplest of these is a single-photon detection, often used for DV QIP.

\begin{figure*}[!ht]
  \includegraphics[width=.88\textwidth]{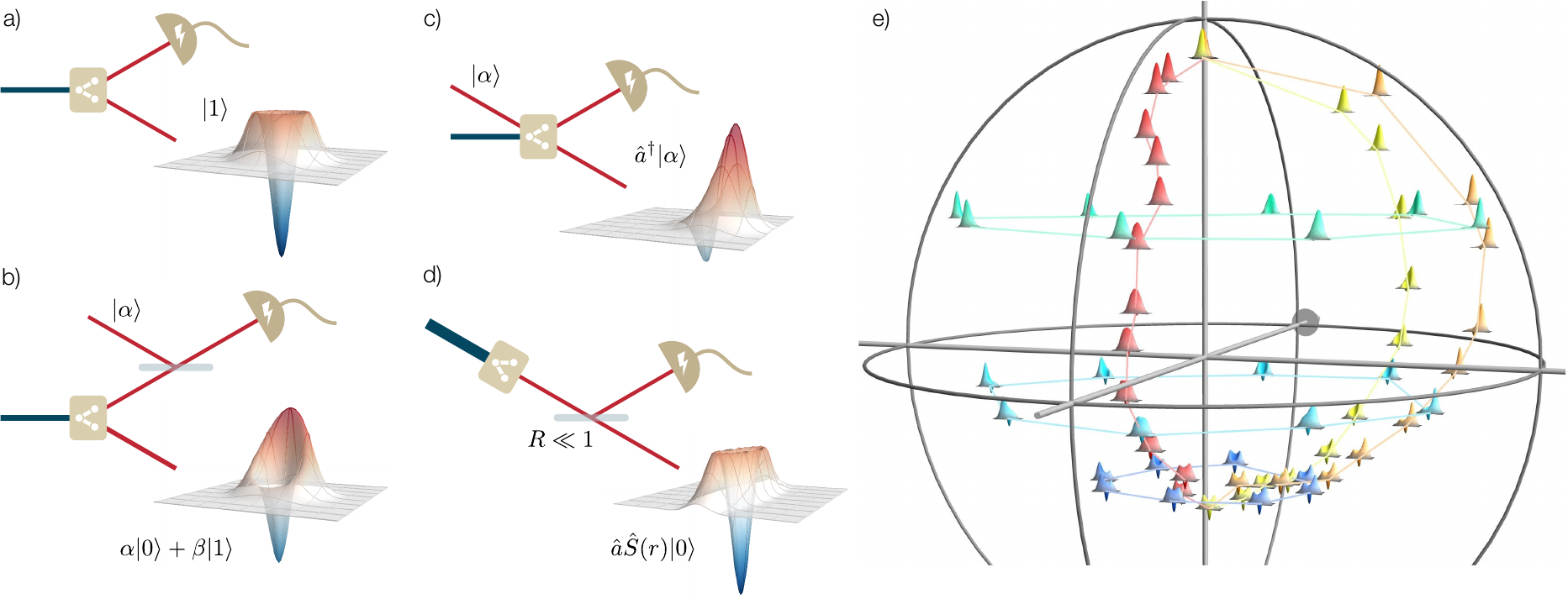}
  \caption{
    Examples of heralded non-Gaussian state preparation schemes that have been
    demonstrated and characterized by quantum homodyne tomography. a) Generation of a single-photon state by SPDC followed by single photon detection b) Engineering of a qubit via displacement and single photon detection c) Addition of a single photon by driven PDC followed by photon detection d) Generation of a kitten state by subtracting a single photon from a squeezed vacuum state e) Wigner functions of various kitten state qubits illustrated on the qubit Bloch sphere \cite{Neergaard-Nielsen2010}.
  }
  \label{fig:ng_gen}
\end{figure*}

An essential workhorse for both DV and CV optical QIP is the parametric amplifier (OPA). The OPA operates by the process of spontaneous parametric down-conversion where a single pump photon is converted into two lower-frequency photons, called signal and idler, in a nonlinear medium.
As the process is spontaneous the probability of generating one or more photon pairs is small, as reflected in the Fock basis representation of the OPA output, 
\begin{equation}
\label{eq:opa}
|\phi\rangle \propto |0_s0_i\rangle + \sqrt{\lambda}|1_s1_i\rangle + \lambda|2_s2_i\rangle + \ldots,
\end{equation}
with $\lambda$ related to the gain of the amplifier. 
It is clear that the signal (s) and idler (i) modes are photon number correlated, but despite of this, the state is Gaussian and thus not directly suitable for various QIP tasks.
However, by combining the CV Gaussian entanglement source with a DV detector it is possible to probabilistically induce the generation of a non-Gaussian state. The detection of a photon in the idler mode of a low-gain OPA heralds the existence of its signal twin (see Fig.~\ref{fig:ng_gen}a). The heralding means that this photon can be rigorously characterized by CV 
homodyne tomography \cite{Lvovsky2001,Neergaard-Nielsen2007,Morin2012}.
By using photon-number-resolving detection, higher photon number Fock states can be generated \cite{Ourjoumtsev2006a,Cooper2012}.
The signal mode of the OPA can be seeded with a Gaussian state, like a thermal or coherent beam, as illustrated in Fig.~\ref{fig:ng_gen}c. The detection of the idler photon then heralds a photon added version of the seeded state, $|\psi_\mathrm{herald}\rangle \propto \hat{a}^\dagger |\psi_\mathrm{seed}\rangle$, which has now turned non-Gaussian \cite{Zavatta2004}.

An experimentally easier operation is the reverse process, namely photon subtraction. This can be implemented simply
by tapping off a small portion of the beam on a beam splitter with
reflectivity $R\ll 1$ and monitoring the reflected part with a photon
detector (see Fig.~\ref{fig:ng_gen}d). The detection of a photon heralds the subtraction of
a photon from the state in the transmitted beam, $|\psi_\mathrm{out}\rangle \propto \hat{a}|\psi_\mathrm{in}\rangle$  \cite{Ban1994}.
Coherent states are eigenstates of the annihilation operator, so photon subtraction has no effect on coherent and thermal states (which are coherent state mixtures). With other initial states, though, photon subtraction acts as a ``de-Gaussifying'' operation. 
Prominently, this has been shown in several experiments with single-mode squeezed vacuum as input (the state in \eqref{eq:opa} with degenerate $s$ and $i$ modes) with subtraction of a single photon \cite{Wenger2004, Ourjoumtsev2006, Neergaard-Nielsen2006, Wakui2007} and later two or three photons \cite{Takahashi2008,Gerrits2010,Namekata2010}. 
The considerable interest in the photon-subtracted squeezed vacuum
stems from the recognition \cite{Dakna1997} that such states are
close-to-ideal approximations to a superposition of 
coherent states (a ``Schr\"odinger's cat'') with small amplitudes, also known as a kitten state.

With just a single photon detector to provide a non-Gaussian element, CV Gaussian operations like phase-space displacement and homodyne detection can be added to the mix to increase the variety of state generation and manipulation. The displacement operation is easily performed by mixing with a strong coherent state on a highly imbalanced beam splitter.
If, as shown in Fig.~\ref{fig:ng_gen}b, a displacement is performed on the idler mode in a heralded single photon setup, one can prepare an arbitrary superposition of the vacuum and the single photon state -- a single-rail photonic qubit, $c_0\ket{0} + c_1\ket{1}$ -- with the coefficients controlled by the amplitude and phase of the displacement \cite{Lvovsky2002,Resch2002,Babichev2003}. This can of course be extended to higher photon numbers by increasing the number of detectors \cite{Bimbard2010,Yukawa2013} as in Fig. \ref{fig:systems}d.
Similarly, displacement before photon subtraction in a kitten state generation setup allows for arbitrary superpositions of the odd and even states \cite{Neergaard-Nielsen2010}. Experimental examples of such superpositions are shown on the Bloch sphere in Fig.~\ref{fig:ng_gen}e.
The amount of control that can be achieved by these simple means is such that any single-mode quantum state can be generated by successive applications of phase-space displacement and either photon addition or subtraction \cite{Dakna1999,Fiurasek2005}.
Even more flexibility would be obtained by employing a photon-number resolving detector in combination with the displacement. Such a hybrid detector was characterized by full tomography in Ref. \cite{Zhang2012}.
As an alternative to displacement, a homodyne detection on one mode of a two-mode state can also induce arbitrary superposition states like the single-rail qubits conditionally \cite{Babichev2004a}. The homodyne measurement outcome then determines the coefficients.

By combining these hybrid techniques, the possibilities of quantum state engineering and information processing become countless. Several schemes for increasing the amplitude of kitten states have been proposed, for example by conditional photon counting or homodyne detection on a combination of multiple kittens \cite{Lund2004, Takeoka2007,Laghaout2013}, displacement-improved photon subtraction \cite{Nielsen2007,Yukawa2013}, temporally separated two-photon subtraction \cite{Takeoka2008}, or conditional squeezing of a Fock state by homodyne projection~\cite{Ourjoumtsev2007a}, see Fig. \ref{fig:systems}e. The energy of one mode of a single-photon entangled state can be increased by displacement~\cite{Bruno2013,Lvovsky2013}. Entangled coherent states can be generated by non-local photon subtraction over a lossy channel~\cite{Ourjoumtsev2009}, and these entangled cat states can be made arbitrarily large through conditional homodyne detection~\cite{Brask2010}. 
Finally, different ways of inducing hybrid entanglement of cat states or coherent states with a microscopic degree of freedom -- a manifestation of Sch\"odinger's cat--atom entanglement -- have recently been proposed~\cite{Ghobadi2013,Andersen2013b} and experimentally demonstrated~\cite{Jeong2014,Morin2014}.

The probabilistic execution of a non-Gaussian transformation of Gaussian states by a non-Gaussian measurement has also been proposed for other systems. This includes the formation of non-Gaussian spin states \cite{Christensen2013,McConnell2013}  as well as non-Gaussian states of mechanical oscillators \cite{Paternostro2011,Galland2014} by means of photon counting detections. These schemes are however experimentally challenging and have not yet been realized, although some progress towards a heralded non-Gaussian spin state has been made \cite{Christensen2013}.

\section{Quantum information processing}
\label{sec:qip}

In the preceding section, we described how hybridization between DV and CV devices and states can be exploited to engineer exotic, non-classical, non-Gaussian quantum states. We shall now discuss possible schemes for quantum information processing that incorporate the above-mentioned hybrid techniques. These include the fundamental tasks of quantum teleportation, quantum error correction (detection), and entanglement distillation, as well as testing Bell inequalities and performing Bell measurements. Ultimately, they would aim at the realization of scalable quantum communication and universal, fault-tolerant quantum computation.

\begin{figure}[t]
  \includegraphics[width=.96\columnwidth]{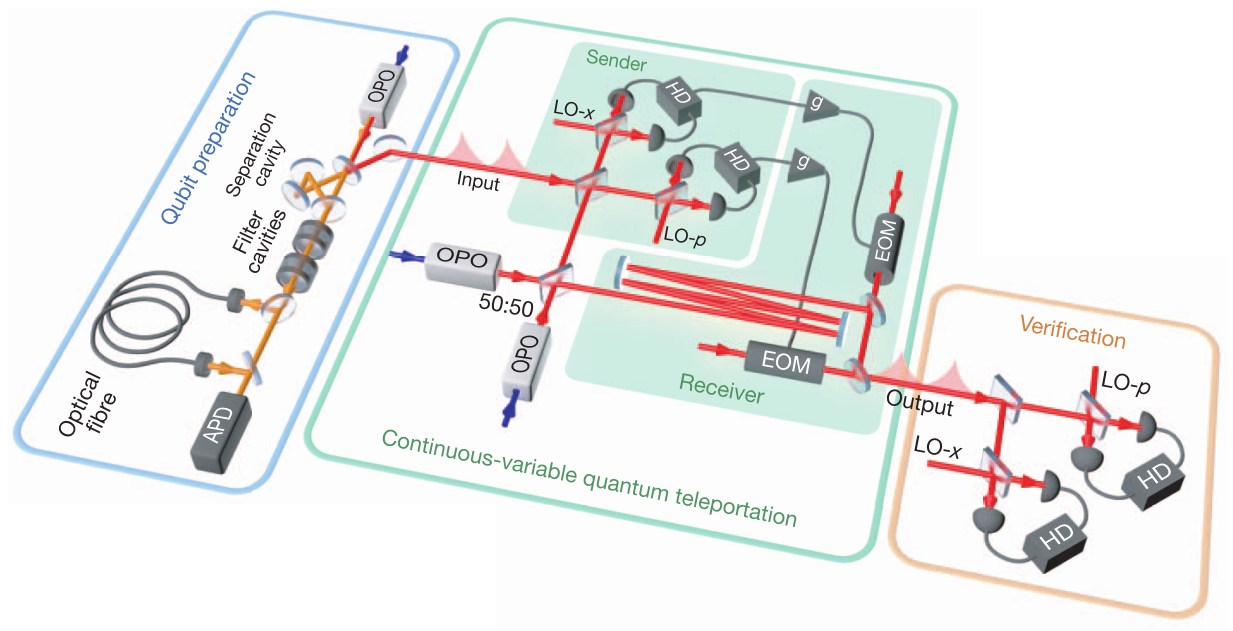}
  \caption{
  Schematic layout of CV teleportation of a DV qubit, as implemented in Ref. \citep{Takeda2013}.
  At \textbf{left}, a time-bin qubit is prepared. The low-gain OPO (optical parametric oscillator) spontaneously emits signal and idler photons which in this case are separated based on their different frequencies.
The idler photon traverses an imbalanced Mach-Zender interferometer after which it is observed by a photon detector (APD, avalanche photo diode). Since the two possible paths (short or long) through the interferometer are indistinguishable, the detection of the idler heralds the signal photon (labelled ``input'') in a qubit state -- a superposition of an early ($\ket{0}$) and a late ($\ket{1}$) time-bin \cite{Zavatta2006}.
  In the \textbf{middle} is the CV teleportation device. Mixing the outputs of two high-gain OPOs results in a two-mode squeezed state. One part of this entangled state is mixed with the input qubit followed by a CV Bell measurement consisting of $x$/$p$ homodyne detectors (HD), while the other part is sent to the receiving station. Here it is modulated with electro-optic modulators (EOM) by the Bell detector outputs amplified with a variable gain. This results -- deterministically -- in a teleported output qubit state almost identical to the input as verified by the homodyne detection at the \textbf{right}.
  }
  \label{fig:teleportation}
\end{figure}

{\bf Hybrid quantum teleportation}. The most elementary protocol in quantum communication is quantum teleportation \cite{Bennett1993} -- the reliable transfer of arbitrary quantum states using shared entanglement and classical communication; and the most obvious hybrid approach to quantum teleportation is CV quantum teleportation \cite{Braunstein1998} of DV states or DV quantum teleportation of CV states. In the optical domain, the former can be, in principle, straightforwardly applied upon any quantum states including single-photon-based qubits. This teleporter possesses the great advantage of being deterministic with solely linear components. However, the price to pay is the intrinsically limited performance: Perfectly faithful and deterministic teleportation of an arbitrary state can only be attained in the limit of an unphysical, infinite degree of Gaussian entanglement. 
Deterministic CV teleportation of DV states has recently been demonstrated on photonic qubits \cite{Takeda2013} and also for a cat state \cite{Lee2011} (see Fig. \ref{fig:teleportation}).

The converse quantum teleporter, using DV entanglement and DV operations to transfer a CV state, requires breaking up a high-dimensional CV state into states of smaller dimension and performing correspondingly many individual DV teleportations \cite{Andersen2013a,Marshall2014,Kogias2014}. In contrast to the standard CV teleporter, the optical DV teleporter can reach fidelities of 100\%. However, its efficiency is fundamentally limited by the probabilistic nature of qubit Bell measurements with linear transformations \cite{Calsamiglia2001}. Only by the use of non-Gaussian transformations or non-Gaussian ancillary states can the teleporter become (near-)deterministic.

Hence, the example of quantum teleportation illustrates nicely what an optical hybrid approach does: it can turn an otherwise probabilistic, linear-optical qubit teleporter into a fully deterministic device, possibly at the expense of the transfer fidelity; and it can make use of a potentially high-fidelity transfer of low-dimensional optical states for reliably transmitting a CV state, at the expense of a non-unit success probability. This new level of versatility is, of course, even greater when matter systems are included, as the light-matter interactions offer an alternative way of performing efficient Bell measurements. In fact, using atomic ensembles or two-level emitters, such hybrid light-matter teleportations have been already proposed for long-distance quantum communication \cite{Duan2001} and, on a small scale, experimentally demonstrated \cite{Sherson2006,Yuan2008}. 

Cat-state qubits (defined as $a|\alpha\rangle + b|-\alpha\rangle$ where $a,b$ are complex numbers) may also be teleported using an entangled state of two cat-qubits such as $|\alpha,\alpha\rangle + |-\alpha,-\alpha\rangle$. This protocol has been experimentally demonstrated for binary coherent states \cite{Neergaard-Nielsen2013}, and it forms the critical element of cat-qubit quantum computing \cite{Jeong2002,Ralph2003,Lund2008} of which a few probabilistic gates have been realised \cite{Tipsmark2011,Blandino2012}. It is also interesting to note that the efficient distribution of entangled cat states for cat state teleportation can be achieved using a another hybrid approach that combines DV and CV measurements~\cite{Brask2010}.

{\bf Hybrid entanglement distillation and quantum communication}. For quantum communication based upon the distribution of entangled states, like in a quantum repeater, it is desirable to initially prepare and distribute optical entanglement with high efficiency. Since the CV Gaussian entangled states can be produced in an unconditional fashion, they may serve as a deterministic source of shared entanglement, thus saving expensive storage times in a fully fledged quantum repeater. However, Gaussian entanglement is very sensitive to photon losses and hence entanglement distillation will be absolutely necessary. Solely using CV Gaussian operations does not allow for distilling high-quality Gaussian entanglement from low-quality, noisy Gaussian entanglement \cite{Eisert2002,Fiurasek2002,Giedke2002}. The required non-Gaussian element may then be introduced through photon subtraction. For the pure, photon-number correlated, two-mode squeezed state, local photon subtractions acting on the two-mode number states as $(\hat a \otimes \hat a) |n,n\rangle =n |n-1,n-1\rangle$ would smoothen the photon-number distribution, effectively enhancing the entanglement of the state \cite{Opatrny2000}. Such an enhancement can also be obtained when the initial states are mixed and noisy after a lossy channel transmission \cite{Kitagawa2006,Zhang2010}. Experiments were already performed, showing such photon subtraction-based distillation \cite{Ourjoumtsev2007,Takahashi2010}.

Apart from photon subtraction, photon addition can be a useful DV operation to locally improve the entanglement of bipartite Gaussian states \cite{Navarrete-Benlloch2012}. In fact, coherent linear combinations of photon subtraction $\propto \hat a$ and addition $\propto \hat a^{\dagger}$ may be the optimal choice \cite{Zavatta2004}; and one way to obtain such superpositions is once again a hybrid technique: combine DV photon subtraction with a CV squeezing operation, $\hat S^{\dagger}(r) \hat a \hat S(r) =\cosh r \hat a + \sinh r \hat a^{\dagger}$ where $r$ is the squeezing parameter. Notice that in this case, squeezing would be promoted from a sole offline resource to an online tool performed locally and individually on the two halves of a two-mode CV Gaussian state.

As squeezing, in conjunction with photon subtraction, allows for obtaining new kinds of hybrid operations -- adding and subtracting light particles in a coherent, wave-like fashion -- it opens up completely new possibilities for quantum communication, for instance, by optimizing entanglement distillation schemes \cite{Zhang2011}. Remarkably, even simpler hybrid operations such as combinations of CV phase-space displacements with DV photon subtractions may still allow for such improvements \cite{Fiurasek2011,Tipsmark2013}. The effect of these entanglement-enhancing operations is to non-locally transform the initial CV Gaussian states into, in lowest order, DV qubit-type entangled states -- a kind of bipartite hybrid state engineering at a distance.

Besides supplementing photon subtraction by squeezing or displacement operations, there are other recently proposed techniques that allow for optically realizing entanglement distillation. One such prominent method is the heralded, noiseless linear amplifier (NLA) introduced by Lund and Ralph \cite{Ralph2009} and realized in different ways \cite{Xiang2010,Ferreyrol2010,Zavatta2010}. On the operational, application-oriented side, the NLA provides a DV, non-Gaussian tool to distill CV \cite{Ralph2009} as well as DV \cite{Minar2012} entangled states, where in the latter case, this may immediately render atomic-ensemble-based quantum repeaters \cite{Duan2001} more efficient \cite{Sangouard2011}. Moreover, the DV NLA can be used to correct errors in CV teleportation resulting from a lossy distribution of entangled states \cite{Ralph2010}. On the more conceptual side, the NLA serves as yet another illustrative example for the benefit of a hybrid approach: while the original, non-heralded and deterministic, quantum-limited (phase-insensitive) linear amplifier by Caves \cite{Caves1981} is a CV Gaussian operation, only its generalization to the non-Gaussian and heralded DV regime enables one to beat supposedly fundamental quantum noise limits, again at the expense of a non-unit success probability.


{\bf Hybrid Bell tests and measurements}. The list of both fundamental and application-oriented quantum tasks and protocols in a quantum communication setting, i.e., with at least two spatially separated parties sharing classical and quantum resources, can be further extended. There are the fundamental Bell tests, which are to negate local realism even by means of realistic, imperfect detectors, and for this, combined DV and CV measurements may help \cite{Cavalcanti2011,Laghaout2011}. Gaussian CV entangled states, having an immediate hidden-variable description in form of a positive Wigner function, can never violate a Bell inequality based upon CV homodyne measurements; only the detection of a set of discrete variables such as photon-number parities \cite{Banaszek1998} would work. In a quantum repeater chain based upon elementary, polarization-entangled photon pairs, the necessary two-qubit Bell measurements for connecting the individual repeater segments can be made more efficient using -- once again -- local squeezers, in addition to counting photon numbers \cite{Zaidi2013}.

\begin{figure}[!ht]
  \includegraphics[width=.95\columnwidth]{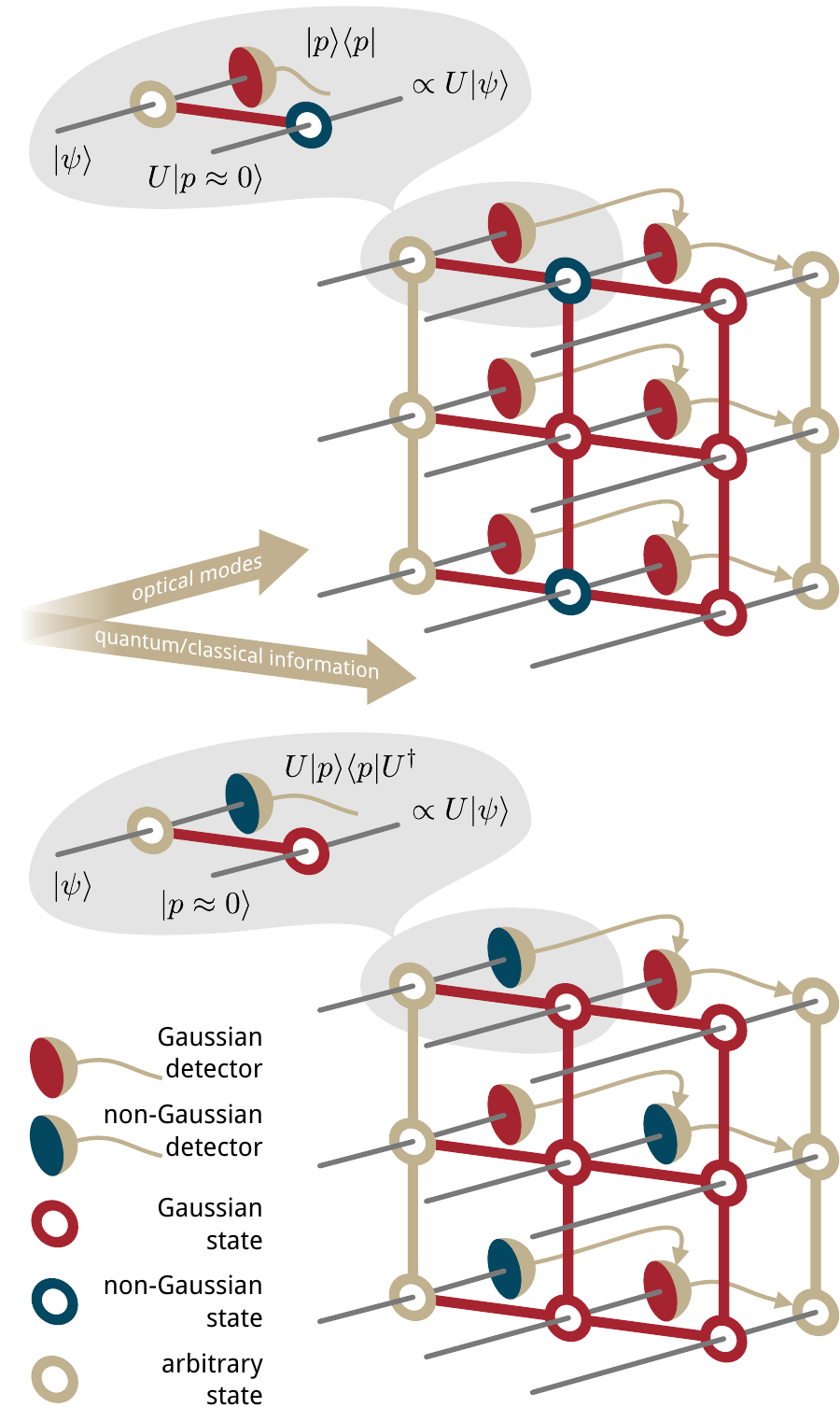}
  \caption{
  Measurement-based quantum computation using two-dimensional lattices
  corresponding to offline-prepared, optical multi-mode cluster states.
  The lattices are built from single-mode states through Gaussian two-mode
  interactions (squeezers and beam splitters; thick red lines).
  Arbitrary multi-mode optical states (figure: three modes; vertically oriented
  input and output modes in gold) can be processed by individually measuring
  the propagating optical modes (thin grey lines) except for the output and
  feedforwarding the measurement results. Quantum and classical (feedforward)
  information evolve from left to right.
  Top: some of the Gaussian squeezed single-mode states (red) of the cluster are replaced
  by non-Gaussian single-mode states (blue); all measurements are Gaussian homodyne detections (red).
  Bottom: some of the Gaussian detectors (red) are replaced by non-Gaussian detectors
  (e.g. photon counters; blue); all initial single-mode states are Gaussian squeezed states (red),
  and hence the entire cluster state is Gaussian. Universal operations, i.e., arbitrary
  output states, can be achieved either way, through CV measurements on non-Gaussian states
  or arbitrary measurements on Gaussian states. For arbitrarily long computations, 
  the accumulation of errors caused
  e.g. by finite squeezing must be suppressed via some form of quantum error correction.  
  }
  \label{fig:computation}
\end{figure}

{\bf Hybrid quantum computing}. While a full-scale quantum repeater can be thought of as a nonlocal device with local, small-scale quantum computers that benefits from the hybrid approach for quantum error detection and entanglement distillation, a large-scale, universal quantum computer would have to combine the universal processing of some logical quantum state with a sufficient degree of fault tolerance in order to maintain an arbitrarily long computation. Examples of universal gate sets leading to universal computation for DV and CV logical encoding are given in BOX 1 and 2.

The optical hybrid approach to implementing universal quantum gates \cite{VanLoock2011,Furusawa2011} relies on teleportation-based, or, more generally, measurement-based quantum processing \cite{Raussendorf2001}, similar to Ref.~\cite{Knill2001}. However, the crucial and with regards to implementations useful difference is that some parts of an otherwise fully DV (CV) measurement-based scheme are replaced by CV (DV) elements.

Consider the elementary circuits in the zoomed grey circles of Fig.~\ref{fig:computation}: an arbitrary input quantum state $|\psi\rangle$, representing one propagating optical mode (indicated by the corresponding thin grey line), gets entangled with either a Gaussian, highly squeezed single-mode state $|p\approx 0\rangle$ (indicated by a red circle) or, alternatively, with a suitable non-Gaussian single-mode state $U\,|p\approx 0\rangle$ (indicated by a blue circle). The entangling operation is Gaussian in either case (indicated by the thick red line). A universally transformed single-mode output state, $U\,|\psi \rangle$, can then be produced, either through simple Gaussian homodyne detection projecting the input mode onto the $|p\rangle$ basis (indicated by a red detector) or via a, in general, non-Gaussian projection measurement onto a suitably rotated measurement basis, $U\,|p\rangle$ (indicated by a blue detector). These elementary processes are reminiscent of CV quantum teleportation, where indeed the final, desired output state would be obtained only depending on some phase-space displacement given by the outcome of the measurement. The application of an arbitrary single-mode gate $U$ may require more than just one such elementary step, where generally every step includes adjusting the measurement basis conditioned upon the previous outcomes and depending on some additional corrections beyond simple displacements. The crucial difference compared with standard quantum teleportation is (besides the measurement being only single-mode instead of a two-mode Bell measurement) that an input state is not only transferred, but also manipulated.

Using these elementary schemes, one may realize two important gates of the CV universal gate set: a universal squeezer \cite{Filip2005} (experimentally applied to a single photon in Ref. \cite{Miwa2014}) and a cubic phase gate \cite{Gottesman2001}. For the case, when the ancilla state is non-Gaussian ($U\,|p\approx 0\rangle$) and the desired gate is weak and, for instance, cubic, a suitable three-photon superposition state (as demonstrated in Ref. \cite{Yukawa2013}) may be employed \cite{Marek2011}.

In order to achieve universal processing of arbitrary multi-mode quantum states, the elementary teleportation circuits can be concatenated in a so-called cluster-state computation \cite{Raussendorf2001,Menicucci2006}. The input state is then teleported into (or prepared within) a sufficiently large, multi-mode, non-Gaussian (or Gaussian) cluster state, which is then subject to CV Gaussian (or arbitrary, including DV non-Gaussian) measurements (see Fig.~\ref{fig:computation}). While universality does not depend on whether the non-Gaussian elements appear in the detectors or in the ancilla states, the choice of hybridization will often be based on the actual complexity of the implementation.

Similar to quantum communication, hybrid schemes for quantum computation also greatly benefit from combining light and matter systems. Again, the additional possibility of coupling light with matter, that is more specifically, CV oscillators with DV two-level systems, extends the toolbox of gate operations to include highly nonlinear interactions. This allows for deterministic universal gate operations on the qubits \cite{Wang2002}, where again the light mode acts as a quantum bus that mediates the qubit gates. However, this time the light mode is no longer measured out, but instead it disentangles automatically from the qubit systems when a full sequence of light-matter interactions has been completed. Suitable interactions for this purpose are the controlled phase rotations induced by dispersive, cavity QED-type atom-light interactions \cite{VanLoock2008} or the naturally occurring, controlled phase-space displacements of a microwave mode depending on the discrete state of a superconducting qubit \cite{Spiller2006}. It is important to notice that in these schemes, in principle, weak nonlinear interactions are sufficient, which result in small phase shifts or displacements that can be effectively enhanced by increasing the amplitude of the CV quantum bus.

\section{Outlook}
\label{sec:outlook}

Less than a decade ago, the boundary between DV and CV platforms for quantum information processing was
extremely sharp. This is no longer the case. The boundary is progressively becoming smeared out as a result of several
recent advances in combining the technologies of the two platforms. This marriage of the two different disciplines, as reviewed in this article, has led to a wealth of new theoretical proposals and some experimental implementations of new promising protocols for quantum information processing. However, the field is still very young and researchers might only have scratched the surface of a much larger and richer field. 

Most of the demonstrations to date are proof-of-principle experiments with a lack of high-fidelity operation, efficiency and scalability. To advance the field in the direction of higher fidelity operation and higher efficiency, a deeper understanding of the present limitations must be attained in order to devise new systems with fidelities and efficiencies that are sufficiently high to allow for perfect error correction and thus fault tolerant QIP. To facilitate scalability, the systems must be miniaturized: E.g. using integrated photonics, phononics and electronics for squeezed light generation and propagation; solid state platforms containing finite-level systems for single photon generation and non-Gaussian transformations; and integrated on-chip detector technology for efficient detection of either CVs and DVs combined with real-time feedback.

\bibliographystyle{naturemag}
\bibliography{hybrid}


\pagebreak

\pagebreak

\end{document}